\documentclass[journal]{IEEEtai}
\usepackage[colorlinks,urlcolor=blue,linkcolor=blue,citecolor=blue]{hyperref}
\usepackage{graphicx}

\usepackage{amsmath, amssymb, amsfonts}
\usepackage{url}
\usepackage{cite}
\usepackage{bm}
\usepackage{arydshln}
\usepackage{algorithmic}
\usepackage{algorithm}
\usepackage{authblk}

\newcommand{\m}{\mathrm}
\newcommand{\p}{\prime}
\newcommand{\argmin}{\mathop{\rm argmin}\limits}
\newcommand{\argmax}{\mathop{\rm argmax}\limits}

\begin{document}
\title{FG-SSA: Features Gradient-based Signals Selection Algorithm of Linear Complexity for \\ Convolutional Neural Networks} 

\author{Yuto Omae, \IEEEmembership{Member, IEEE}, 
Yusuke Sakai, and Hirotaka Takahashi
\thanks{This work was supported in part by JSPS Grant-in-Aid for Young Scientists (Grant No. 19K20062).}
\thanks{Y. Omae is with the Artificial Intelligence Research Center, College of Industrial Technology, Nihon University, Chiba, 275-8575, Japan (e-mail: oomae.yuuto@nihon-u.ac.jp).}
\thanks{Y. Sakai and H. Takahashi are with the Research Center for Space Science, Advanced Research Laboratories, Tokyo City University, Tokyo 158-8557, Japan (e-mail: hirotaka@tcu.ac.jp).}
}

\markboth{This is the preprint version (Feb. 2023).}
{Y. Omae \MakeLowercase{\textit{et al.}}: FG-SSA: Features Gradient-based Signals Selection Algorithm for Convolutional Neural Networks}

\maketitle

\begin{abstract}
Recently, many convolutional neural networks (CNNs) for classification by time domain data of multisignals have been developed.
Although some signals are important for correct classification, others are not.
When data that do not include important signals for classification are taken as the CNN input layer, the calculation, memory, and data collection costs increase.
Therefore, identifying and eliminating nonimportant signals from the input layer are important.
In this study, we proposed features gradient-based signals selection algorithm (FG-SSA), which can be used for finding and removing nonimportant signals for classification by utilizing features gradient obtained by the calculation process of grad-CAM.
When we define $n_\m{s}$ as the number of signals, the computational complexity of the proposed algorithm is linear time $\mathcal{O}(n_\m{s})$, that is, it has a low calculation cost.
We verified the effectiveness of the algorithm using the OPPORTUNITY Activity Recognition dataset, which is an open dataset comprising acceleration signals of human activities.
In addition, we checked the average 6.55 signals from a total of 15 acceleration signals (five triaxial sensors) that were removed by FG-SSA while maintaining high generalization scores of classification.
Therefore, the proposed algorithm FG-SSA has an effect on finding and removing signals that are not important for CNN-based classification. 
\end{abstract}

\begin{IEEEImpStatement}
Recently, it has become possible to measure various signals, e.g., electroencephalogram, electrocardiogram, acceleration/gyro, ambient signals, and others. This has led to the development of many artificial intelligences for classification using multiple signals as input data. However, some of the multiple signals are not necessary for class estimation. The such signals increase memory resources, computation time, data measurement cost, and using artificial intelligence cost. Therefore, it is better to remove signals that are not necessary for class estimation from the input layer of the artificial intelligence. Therefore, in this study, we devised an algorithm to remove them. By applying this algorithm to a artificial intelligence that performs class estimation from multiple signals, useless signals are removed and a lightweight model can be constructed.
\end{IEEEImpStatement}


\section{Introduction}
Recently, many convolutional neural networks (CNNs) have been developed for classification in the time domain of signals \cite{time_cnn_1, time_cnn_2, time_cnn_3, time_cnn_4}. 
Although these studies used all signals, signals that are not important for correct classification may be included in the CNN input layer.
These signals worsen classification accuracy, and increase calculation costs, required memory, and data collection costs.
Therefore, finding and removing nonimportant signals from the CNN input layer and creating a classification model for the minimum signals possible are crucial.
However, many CNNs use all signals \cite{all_sig_1, all_sig_2, time_cnn_4} or manually select signals \cite{manu_sig_1, manu_sig_2}.

We can visually find nonimportant signals using grad-CAM embedded in CNNs.
Grad-CAM \cite{grad_cam_origin} is a method for determining the activated region on input data, and it was proposed by improving the CAM \cite{Zhou_2016_CVPR}.
The method is primarily used to input image data \cite{kara2021covid, kavitha2019multi, nam2021automatic, matsumoto2020diagnosing, hirata2021deep}.
Cases of applying grad-CAM to CNNs that input time domain data of signals are continuously increasing.
For example, 
classification of the sleep stages \cite{sig_grad_2},
prognostication of comatose patients after cardiac arrest \cite{sig_grad_3},
classification of schizophrenia and healthy subjects \cite{sig_grad_4}, and
classification of motor imagery \cite{sig_grad_6}
are performed using electroencephalography (EEG) signal(s).
Electrocardiogram (ECG) signals are used to detect nocturnal hypoglycemia \cite{sig_grad_1} and predict 1-year mortality \cite{sig_grad_5} have been reported.
Acceleration signals have been used to detect hemiplegic gait \cite{acc_grad_1} and human activity recognition \cite{acc_grad_2}.

From these studies, we visually find nonimportant signals for correct classification by applying grad-CAM to CNN, inputting the time domain of the signals.
However, because one grad-CAM for one input data is generated, observing all of them and finding nonimportant signals is extremely difficult.
Therefore, we propose features gradient-based signals selection algorithm (FG-SSA), which can be used to find and remove nonimportant signals from a CNN input layer by utilizing features gradient obtained by the calculation process of grad-CAM.
The algorithm provides a signal subset consisting of only important signals.
When we define $n_\m{s}$ as the number of all signals, 
the computational complexity of the proposed algorithm is linear order $\mathcal{O}(n_\m{s})$, 
that is, FG-SSA has a low calculation cost.

Kim \cite{sig_selec_1} has proposed a group lasso-based algorithm for signal selection.
Although this is an effective method, visual insights into the results are difficult because grad-CAM is not used in the calculation process.
By contrast, by using the proposed algorithm FG-SSA, we can visually determine the reason the removed signals are not important.

\section{Proposed method}
\subsection{Specific problem for classification}\label{se21}
We consider a situation in which the task is to estimate a class $c$ belonging to class set $\bm{C}$ by CNNs from the measurement data of the signal set $\bm{S}$ of size $n_\m{s}$.
Some $n_\m{s}$ signals are important for classification, whereas others are not.
Therefore, finding a signal subset $\bm{S}_\m{use}$ that are removed the nonimportant signals from all signal sets $\bm{S}$ is crucial.
In this study, we provide an algorithm for finding such a subset of signals $\bm{S}_\m{use} \subseteq \bm{S}$.
The applicable targets of the proposed method are all tasks of solving classification problems using CNNs inputting the time domain of multisignals.
For an easy understanding of the principle of the proposed method, we set a specific classification problem and explain the proposed algorithm. 

``OPPORTUNITY Activity Recognition'' is the dataset for activity recognition and it is used in ``Activity Recognition Challenge'' held by IEEE in 2011 \cite{ref_opp1, ref_opp2}.
We regard the dataset as reliable because it is used for the performance evaluation of machine learning in some studies \cite{use_opp1, use_opp2}.
The dataset contains data on multiple inertial measurement units, 3D acceleration sensors, ambient sensors, 3D localization information, etc.
Four subjects performed a natural execution of daily activities.
The activity annotations include locomotions (e.g., sitting, standing, walking) and left- and right-hand actions (e.g., reach, grasp, release, etc.).
Details of the dataset are described in \cite{ref_opp1, ref_opp2}.

In this study, we used five triaxial acceleration sensors from all the measurement devices (sampling frequency: 32 [Hz]).
Attachment points of sensors on the humans body are ``back,'' ``right arm,'' ``left arm,'' ``right shoe,'' and ``left shoe.''
The total number of signals was 15, because we adopted five triaxial sensors ($5 \times 3 = 15$).
We adopted data splitting using the sliding window method with window length $w=60$ and sliding length $60$. 
One signal length was approximately 2 s, because the sampling frequency was 32 [Hz].
Searching for the optimal window length size is important because it is a hyperparameter that affects the estimation accuracy \cite{window_size_imp, window_size_imp2}.
However, because the main objective of this study was to provide a signal-selection algorithm, we did not tune the window length size.

Next, we labeled the motion class label for each dataset based on human activity.
The class labels are combinations of locomotions (``Stand,'' ``Walk,'' ``Sit,'' and ``Lie'') and three hands activity (``R'': moving right hand, ``L'': moving left hand, and ``N'': not moving hands).
For example, the class label ``Sit R'' refers to the sitting and right-hand motions.

Table \ref{tab1} lists the results of applying the described procedures to all the data.
This indicates that the data size of the left-hand motion is small.
We consider that nearly all subjects were right-handed (notably, we could not find a description of the subjects' dominant arm in the explanation of the OPPORTUNITY dataset).
Moreover, data belonging to Lie R and L are absent.
Therefore, these classes were removed from the estimation task;
that is, the total number of classes was 10.

Subsequently, we randomly split all the data into a training dataset (80\%) and a test dataset (20\%).
Moreover, we assigned 20\% of the training dataset to the validation dataset.
The training, validation, and test datasets were independent because we adopted the sliding window method for the same window and slide length (60 steps).

\begin{table}[tb]
  \caption{Samples size of each class label}
  \label{tab1}
  \centering
  \begin{tabular}{crr}\hline
    Labels & Samples & Rates [\%]\\ \hline
    Stand N & 1070 & 20.68\\
    Stand L & 193 & 3.73\\
    Stand R & 642 & 12.41\\ \hdashline
    Walk N & 1405 & 27.16\\
    Walk L & 63 & 1.22\\
    Walk R & 150 &2.90\\  \hdashline
    Sit N & 571 &11.04\\
    Sit L  & 130 &2.51\\
    Sit R & 536 &10.36\\  \hdashline
    Lie N & 414 & 8.00\\ 
    Lie L & 0 & 0.00\\
    Lie R & 0 & 0.00 \\ \hline
    Total & 5174 & 100.00\\
    \hline
  \end{tabular}
\end{table}

 We define the signal set $\bm{S}$, which has 15 elements, and the class set $\bm{C}$, which has 10 elements, as follows:
\begin{align}
\bm{S} = \{&\m{Back\ X}, \m{Back\ Y}, \m{Back\ Z},\nonumber \\ 
&\m{Right\ arm\ X}, \m{Right\ arm\ Y}, \m{Right\ arm\ Z},\nonumber \\
&\m{Left\ arm\ X}, \m{Left\ arm\ Y}, \m{Left\ arm\ Z},\nonumber \\
&\m{Right\ shoe\ X}, \m{Right\ shoe\ Y}, \m{Right\ shoe\ Z},\nonumber \\
&\m{Left\ shoe\ X}, \m{Left\ shoe\ Y}, \m{Left\ shoe\ Z} \}, \label{eq_sigset}
\end{align}
\begin{align}
\bm{C} = \{&\m{Stand\ N}, \m{Stand\ L}, \m{Stand\ R}, \nonumber \\
&\m{Walk\ N}, \m{Walk\ L}, \m{Walk\ R}, \nonumber \\
&\m{Sit\ N}, \m{Sit\ L}, \m{Sit\ R}, \m{Lie\ N}\}.
\end{align}
In other words, CNNs solve 10 classification problems from 15 multisignals.
Moreover, we provide an algorithm for removing nonimportant signals while maintaining the estimation accuracy.
Although the sets $\bm{S}$ and $\bm{C}$ represent acceleration signals and human activities, respectively, the proposed algorithm can be used for other diverse signals, such as EEG, ECG, and others.

\begin{figure*}[t]
 \centering
 \includegraphics[scale=0.23]{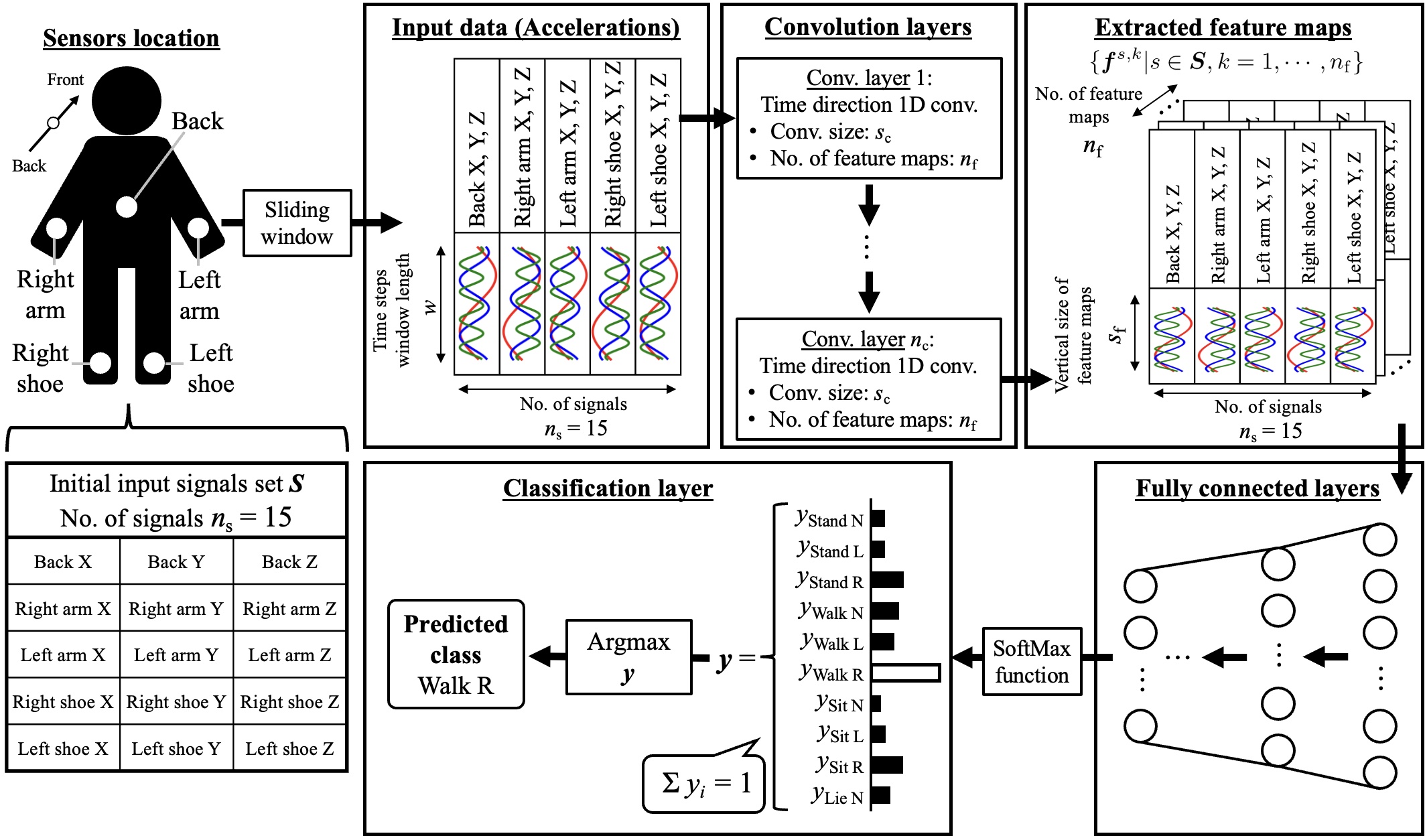}
 \caption{CNN structure for 10 human activities recognition from 15 acceleration signals. We applied the proposed algorithm to this structure.}
 \label{f0}
\end{figure*}

\subsection{Class estimation}
The CNN structure for class estimation is shown in Figure \ref{f0}.
The data for the input layer are in the form of a matrix $w \times n_\m{s}$.
$w$ is the window length and $n_\m{s} = |\bm{S}|$ is the number of signals.
In the case of Subsection \ref{se21}, the input layer size is a matrix of $60 \times 15$.

Subsequently, the input data are convoluted using kernel filters of the time-directional convolution size $s_\m{c}$.
The number of generated feature maps is $n_\m{f}$ because the number of filters is $n_\m{f}$.
The reason for adopting only time-directional convolution is to avoid mixing a signal and others in the convolution process. 
The input data are convoluted by these $n_\m{c}$ convolution layers, and the CNN generates the feature maps shown in the extracted feature maps in Figure \ref{f0}.
Subsequently, the CNN generates the output vector using some fully connected layers and the SoftMax function.
We define the output vector $\bm{y}$ as follows: 
\begin{align}
\bm{y} = [y_c] \in \mathbb{R}^{|\bm{C}|}, \ c \in \bm{C}, \ \sum_{c \in \bm{C}} y_c = 1.
\end{align}
The estimation class $c^{\p}$ is 
\begin{align}
c^{\p} = \argmax_{c \in \bm{C}} \ \{y_c | c \in \bm{C}\}.
\end{align}
Herein, we only define the output vector $\bm{y}$ because it appears in the grad-CAM definition.
We explain grad-CAM for time domain signals in the next subsection.

\subsection{Time-directional grad-CAM}
We define the vertical size of the feature map before the fully connected layer as $s_\m{f}$, as shown in Figure \ref{f0} ``Extracted feature maps.''
Let us denote the $k$th feature map of signal $s$ as 
\begin{align}
&\bm{f}^{s, k} = [f^{s, k}_1 \ \ f^{s, k}_2 \ \ \cdots \ \ f^{s, k}_{s_\m{f}}]^{\top} \in \mathbb{R}^{s_\m{f}}, \nonumber \\
&s \in \bm{S}, \ k \in \{1, \cdots, n_\m{f}\}.
\end{align}
In the case of CNNs for image-data-based classification, the form of feature maps is a matrix.
However, in the case of CNNs consisting of time-directional convolution layers, the form of the feature maps is a vector.
We define the effect of the feature map $\bm{f}^{s, k}$ on the estimation class $c^{\p}$ as 
\begin{align}
\alpha^{s, k}_{c^{\p}} = \frac{1}{s_\m{f}}\sum_{j=1}^{s_\m{f}}\frac{\partial y_{c^{\p}}}{\partial f^{s, k}_j}, \label{eq_alp}
\end{align}
Then, we can apply the grad-CAM of signal $s$ to estimation class $c^{\p}$ in the vector form of 
\begin{align}
\bm{Z}^s_{c^{\p}} = \m{ReLU}\Bigl(\frac{1}{n_\m{f}}\sum_{k=1}^{n_\m{f}} \alpha^{s, k}_{c^{\p}} \bm{f}^{s, k}\Bigr), \ \ \bm{Z}^s_{c^{\p}} \in \mathbb{R}^{s_\m{f}}_{\geq 0}.
\end{align}
The activated region of each signal can be understood by calculating $\bm{Z}^s_{c^{\p}}$ for all signals $\forall s \in \bm{S}$.

In this paper, we refer to $\bm{Z}^s_{c^{\p}}$ as ``time-directional grad-CAM'' because it is calculated by the summing partial differentiations of time direction.
This was defined by a minor change in the basic grad-CAM for image data \cite{grad_cam_origin}.
In the case of the basic grad-CAM \cite{grad_cam_origin}, the CNN generates one grad-CAM for one input data. 
By contrast, in the case of time-directional grad-CAM, CNN generates grad-CAMs as many as the number of signals $n_\m{s}$ for one input data.

\begin{figure*}[t]
 \centering
 \includegraphics[scale=0.55]{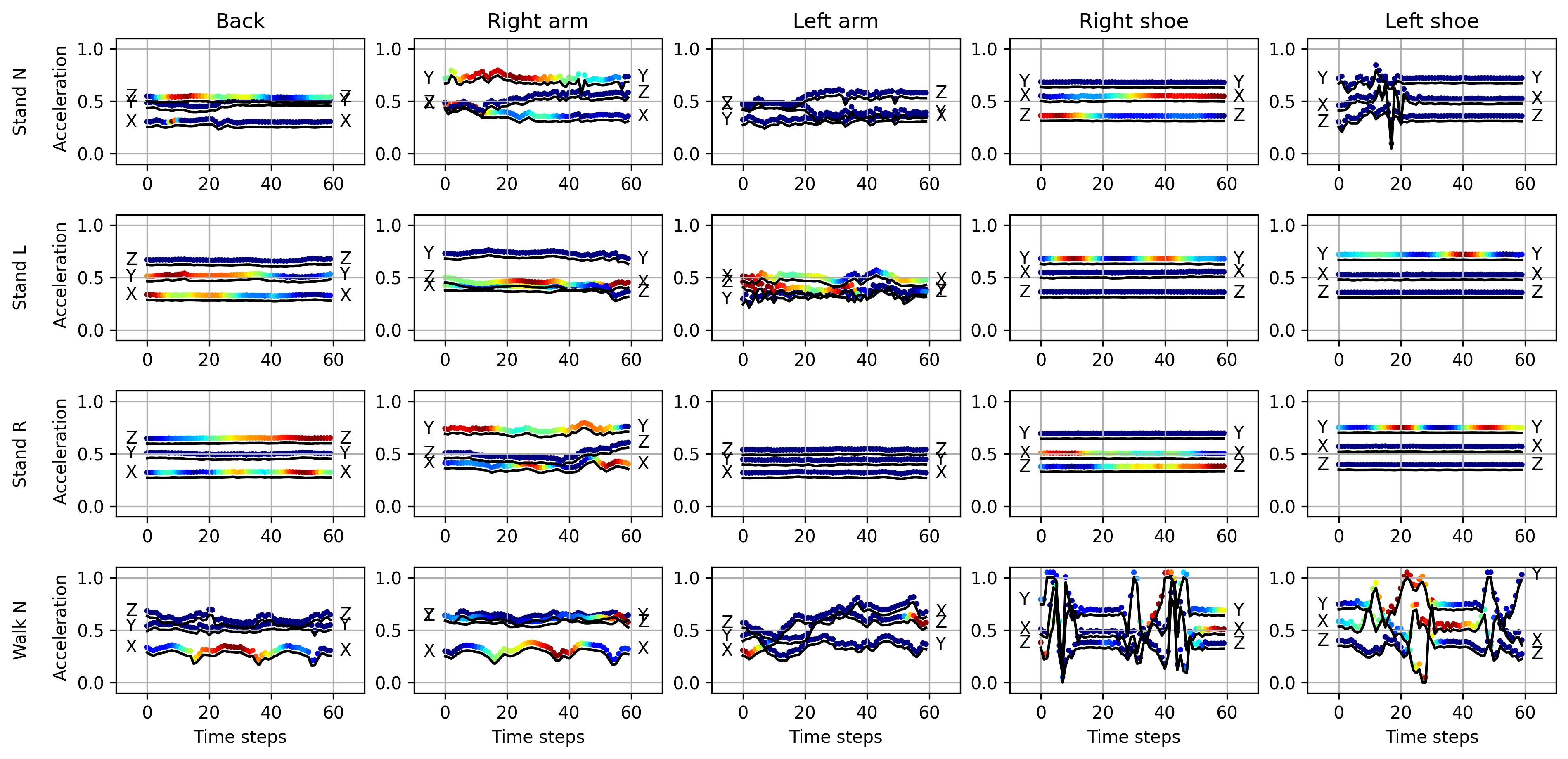}
 \caption{Examples of time-directional grad-CAMs of 15 acceleration signals for four estimation classes (Stand N, Stand L, Stand R, and Walk N) calculated by the CNN of condition A described in Section \ref{sec4}. The red color represents high activation and the dark blue represents low activation.}
 \label{f1}
\end{figure*}

Figure \ref{f1} shows the examples of $\bm{Z}^s_{c^{\p}}$ calculated using the CNN described in Section \ref{sec4}.
From top to bottom, these results correspond to $c^\p = \text{Stand N}, \text{Stand L}, \text{Stand R}$, and $\text{Walk N}$.
We can determine the signals that are important by viewing the time-directional grad-CAM shown in Figure \ref{f1}.
For example, signals from the left arm and left shoe are not used for estimating the Stand N class.
Moreover, left arm signals are not used for the estimation of the Stand R class.
Therefore, the time-directional grad-CAM is effective for finding nonimportant signals.

\subsection{Signals importance index} \label{sec2.4}
Although we can find nonimportant signals by viewing time-directional grad-CAM, the result varies for each input data.
Therefore, when the data size is large, viewing all the grad-CAMs is difficult.
Herein, we quantify the importance of signal $s$ for classification based on $\alpha^{s, k}_{c^{\p}}$ defined in Equation \ref{eq_alp}.

Here, we denote the input dataset of size $n_\text{dat}$ as 
\begin{align}
\bm{X}= \{\bm{X}_i | i=1, \cdots, n_\text{dat}\}, \ \ \bm{X}_i \in \mathbb{R}^{w\times n_\m{s}}.
\end{align}
The size of $i$th input data $\bm{X}_i$ is $w \times n_\m{s}$, which is the window length $w$ and number of signals $n_\m{s}$. 
When we define the input dataset of estimation class $c^\p \in \bm{C}$ as $\bm{X}_{c^\p}$, we can represent the input dataset $\bm{X}$ as
\begin{align}
\bm{X}= \bigcup_{c^\p \in \bm{C}} \bm{X}_{c^\p}.
\end{align}
Using the set $\bm{X}_{c^\p}$, we define the importance of the signal $s \in \bm{S}$ to the estimation class $c^\p \in \bm{C}$ as
\begin{align}
L^{s}(\bm{X}_{c^\p}) = \frac{1}{|\bm{X}_{c^\p}|} \sum_{\bm{X}_i \in \bm{X}_{c^\p}} g^s_{c^{\p}} (\bm{X}_i), \label{}
\end{align}
where 
\begin{align}
&g^s_{c^{\p}}(\bm{X}_i) = \frac{1}{n_\m{f}} \sum_{k=1}^{n_\m{f}} \beta^{s, k}_{c^{\p}}(\bm{X}_i), \nonumber \\
&\beta^{s, k}_{c^{\p}}(\bm{X}_i) = 
\begin{cases}
\alpha^{s, k}_{c^{\p}}(\bm{X}_i), \ \ \text{if }   \alpha^{s, k}_{c^{\p}}(\bm{X}_i) \ge 0 \\
0 , \ \ \text{otherwise}
\end{cases},  \nonumber \\
&\bm{X}_i \in \bm{X}_{c^\p}.
\label{eq_up_zero}
\end{align}
In addition, $\alpha^{s, k}_{c^{\p}}(\bm{X}_i)$ is $\alpha^{s, k}_{c^{\p}}$ of the input data $\bm{X}_i$ to grad-CAM, and $c^{\p}$ is the estimated class.
Therefore, $L^{s}(\bm{X}_{c^\p})$ represents the importance of signal $s$ to class $c^{\p}$ based on the grad-CAM.
Notably, to extract the positive effect on classification, we ignore terms with negative partial derivatives, as shown in Equation \ref{eq_up_zero}.
Moreover, using $L^{s}(\bm{X}_{c^\p})$, we can obtain the matrix 
\begin{align}
&\bm{I}_\m{mat}(\bm{X}) = [L^s(\bm{X}_{c^\p})] \in \mathbb{R}^{|\bm{S}| \times |\bm{C}|}_{\ge 0}\nonumber \\
&s \in \bm{S}, c^\p \in \bm{C}.
\end{align}
We refer to $\bm{I}_\m{mat}(\bm{X})$ as the ``signals importance matrix (SIM)'' because the matrix comprises the importance of all signals and classes using the input dataset $\bm{X}$.
We can understand the effect of each signal to all classes by calculating and viewing SIM $\bm{I}_\m{mat}(\bm{X})$.

\begin{figure*}[t]
 \centering
 \includegraphics[scale=0.7]{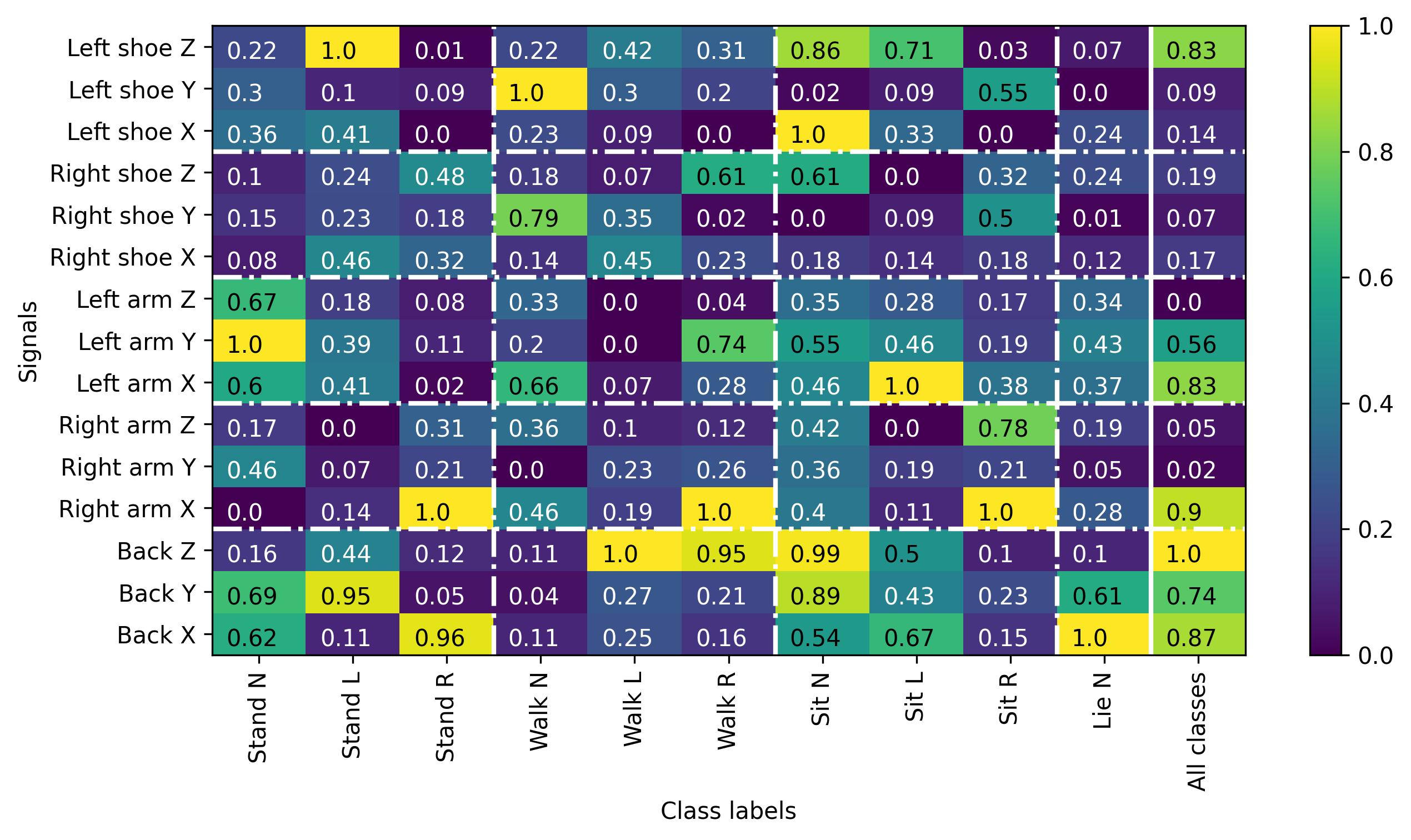}
 \caption{Example of signals importance to all class expressed by SIM $\bm{I}_\m{mat}(\bm{X})$ and SIV $\bm{I}_\m{vec}(\bm{X})$ calculated to CNNs of Condition A described in Section \ref{sec4}. The columns 1 to 10 are SIM and the column 11 is SIV (``All classes''). The values are standardized from zero to one in each column.
 The greater the cell's value, the greater is the relevance of signal $s$ on estimation class $c^\p$.}
 \label{f2.5}
\end{figure*}

Although $\bm{I}_\m{mat}(\bm{X})$ includes important information, summarizing the SIM is necessary to find signals that are not important to all classes.
Therefore, by calculating the summation of the row values of SIM, we denote the importance of signal $s$ as 
\begin{align}
I^s(\bm{X}) = \frac{1}{|\bm{C}|}\sum_{c^\p \in \bm{C}} I^{s, c^\p}_\m{mat}(\bm{X}), \label{eq_imp2}
\end{align}
where $I^{s, c^\p}_\m{mat}(\bm{X})$ is the value of row $s$ and column $c^\p$.
By calculating $I^s(\bm{X})$ for all signals $s$, we obtain the vector as 
\begin{align}
\bm{I}_\m{vec}(\bm{X}) = [I^s(\bm{X})] \in \mathbb{R}^{|\bm{S}|}_{\ge 0}.
\end{align}
We refer to $\bm{I}_\m{vec}(\bm{X})$ as the ``signals importance vector (SIV)'' because it is the vector that consists of the signal importance.

SIM and SIV are shown in Figure \ref{f2.5}.
Examples were calculated using the CNN of condition A described in Section \ref{sec4} (see condition A in Section \ref{sec4}).
The result of SIM (columns 1 to 10 in Figure \ref{f2.5}) includes various important information, for example, the signals Back Z and Right arm X are important for estimating the class Walk R.
Moreover, by viewing SIV (the column 11 in Figure \ref{f2.5}), we can identify the signals that are important for all classes.

From SIV, we denote the minimum important signal $s_\m{min}$ and maximum signal $s_\m{max}$ as 
\begin{align}
(s_\m{min}, \ s_\m{max}) = \Bigl( \argmin_{s \in \bm{S}} {I^s(\bm{X}}), \ \argmax_{s \in \bm{S}} I^s(\bm{X}) \Bigr). \label{eq_imp}
\end{align}
We expect to maintain the estimation accuracy even when removing $s_\m{min}$ because it is a minimum important signal.
By contrast, when we remove $s_\m{max}$ and re-learn the CNN, we expect the accuracy to decrease, because it is the most important signal.

\begin{algorithm}[t]
  \caption{Features gradient-based signals selection algorithm (FG-SSA)}
  \label{alg1}
  \begin{algorithmic}[1]
  \REQUIRE Training and validation dataset $\bm{X}_\m{train}$ and $\bm{X}_\m{valid}$, 
  input signals set $\bm{S}$, \\ maximum number of using signals $\gamma$
  \ENSURE Using signals set $\bm{S}_\m{use}$
  \STATE Initialization of a signals set $\bm{S}_0 \leftarrow \bm{S}$
  \FOR{$t = 0$ to $|\bm{S}| - 1$}
  \STATE Learning a CNN by the training dataset $\bm{X}_\m{train}^{\bm{S}_t}$
  \STATE Calculating validation accuracy $A(\bm{X}_\m{valid}^{\bm{S}_t})$
  \STATE Finding a minimum importance signal $s_\m{min} \leftarrow \argmin_{s \in \bm{S}_t} I^s(\bm{X}_\m{valid}^{\bm{S}_t})$
  \STATE Removing a signal: $\bm{S}_{t+1} \leftarrow \bm{S}_t \setminus s_\m{min}$
  \ENDFOR
  \STATE $\bm{S}_\m{use} \leftarrow \argmax_{\bm{S}^\p \in \{\bm{S}_0, \cdots, \bm{S}_{|\bm{S}|-1}\} } A(\bm{X}_\m{valid}^{\bm{S}^\p}), \ \m{ s.t., } \ |\bm{S}_\m{use}| \le \gamma$
  \RETURN $\bm{S}_\m{use}$
  \end{algorithmic}
  Note: $\bm{X}_\m{train}^{\bm{S}_t}$ and $\bm{X}_\m{valid}^{\bm{S}_t}$ are all the input data belonging to the training and validation dataset using the signal set $\bm{S}_t$, respectively.
\end{algorithm}

\subsection{Signals Selection Algorithm}
In this study, we proposed an algorithm to find a desirable signal subset $\bm{S}_\m{use} \subseteq \bm{S}$ by removing nonimportant signals.
The proposed method is presented in Algorithm \ref{alg1}.
The main inputs to the algorithm are the training dataset $\bm{X}_\m{train}$, validation dataset $\bm{X}_\m{valid}$, and the input signal set $\bm{S}$.
To avoid data leakage, the test dataset was not used in the algorithm.
The elements of set $\bm{S}$ are the signal identification names defined in Equation \ref{eq_sigset}.
The first procedure is to create the initial signal set $\bm{S}_0$, which consists of all signals (line 1).
The next step is to develop a CNN using the training dataset $\bm{X}_\m{train}^{\bm{S}_0}$, which consists of $\bm{S}_0$ (line 3).
Subsequently, we measure the validation accuracy $A(\bm{X}_\m{valid}^{\bm{S}_0})$ using the validation dataset $\bm{X}_\m{valid}^{\bm{S}_0}$ (line 4).
Next, we determine the most important signal $s_\m{min} \in \bm{S}_0$ based on Equation \ref{eq_imp} (line 5).
Finally, we obtain the next signal set $\bm{S}_1$ by removing $s_\m{min}$ from $\bm{S}_0$ (line 6).

This procedure is repeated until the number of signals reaches one (i.e., $t = |\bm{S}| - 1$), and record validation accuracies.
The algorithm returns the signal subset $\bm{S}_{\m{use}}$ leading to maximum validation accuracy (lines 8 and 9).
Notably, the case wherein the signal subset leading to the maximum accuracy is the initial signal set can occur, that is, $\bm{S}_{\m{use}} = \bm{S}$.
In cases wherein the main purpose is to achieve maximum accuracy, adopting the initial signal set as an optimal subset is a better option, if $\bm{S}_{\m{use}} = \bm{S}$ occurs.
However, some cases exist that require a decrease in the number of signals.
Therefore, we prepare constraints wherein the number of adopted signal sizes is $\gamma$ or less, that is, $|\bm{S}_{\m{use}}| \le \gamma$.
This is a hyperparameter of the proposed algorithm.

When the size of the initial signal set $\bm{S}$ is $n_\m{s}$, the number of developing CNN in Algorithm \ref{alg1} is as follows:  
\begin{align}
T(n_\m{s}) &=n_\m{s}-1 \nonumber \\ & \sim n_\m{s}.
\end{align}
That is, the computational complexity of Algorithm \ref{alg1} is $\mathcal{O}(n_\m{s})$.
This means that even if the signal size $n_\m{s}$ increases, the signal subset is returned in realistic time.

Generally, the total pattern for choosing $m$ from $n$ signals is ${}_{n}\m{C}_{m}$.
Because $m$, the number of signals leading to the maximum validation accuracy, is not known, the total number of combinations of input layers $T_\m{bs}(n_\m{s})$ is similar to
\begin{align}
T_\m{bs}(n_\m{s}) &= \sum_{m=1}^{n_\m{s}} {}_{n_\m{s}}\m{C}_{m} \nonumber \\
& \sim \max_{m} {}_{n_\m{s}}\m{C}_{m} \nonumber \\
& = {}_{n_\m{s}}\m{C}_{\lfloor n_\m{s}/2 \rfloor}.
\end{align}
Therefore, finding the optimal signal subset using a brute-force search is difficult.
The computation time tends to be large because the CNN includes many other hyperparameters.
From this viewpoint, a fast algorithm such as the proposed method is important.

\begin{figure*}[t]
 \centering
 \includegraphics[scale=0.7]{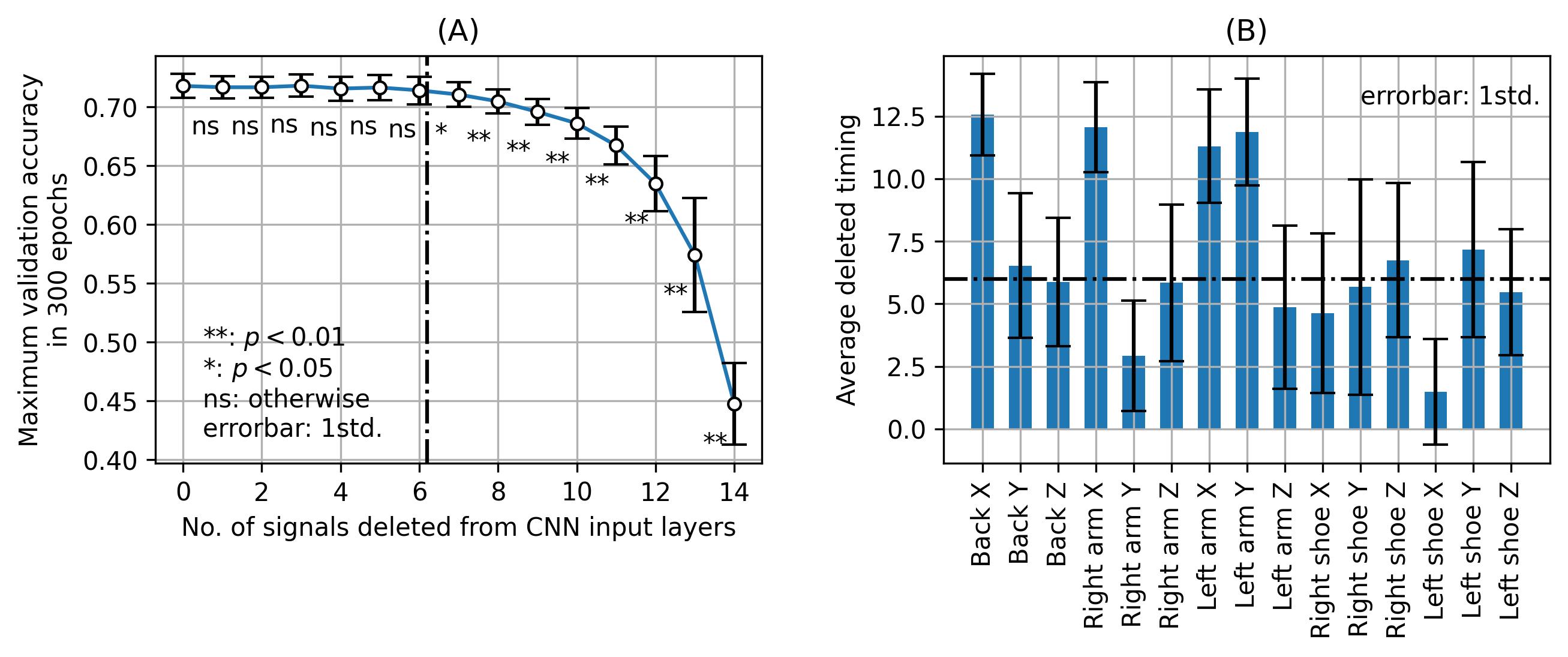}
 \caption{
 (A) Maximum validation accuracy in 300 epochs when the most nonimportant signal $s_\m{min}$ is gradually removed and CNNs are re-learned.
 (B) Average removed timings of each signal.
 It means the smaller the timing, the earlier the signal is removed. 
 The results of both (A) and (B) are average values of 100 seeds, and the error bars are standard deviations.
 The $p$ values in (A) represent the results of two-sided $t$ test, and the dashed line represents the timing of statistically decreasing validation accuracy.}
 \label{f2}
\end{figure*}

\begin{figure*}[t]
 \centering
 \includegraphics[scale=0.7]{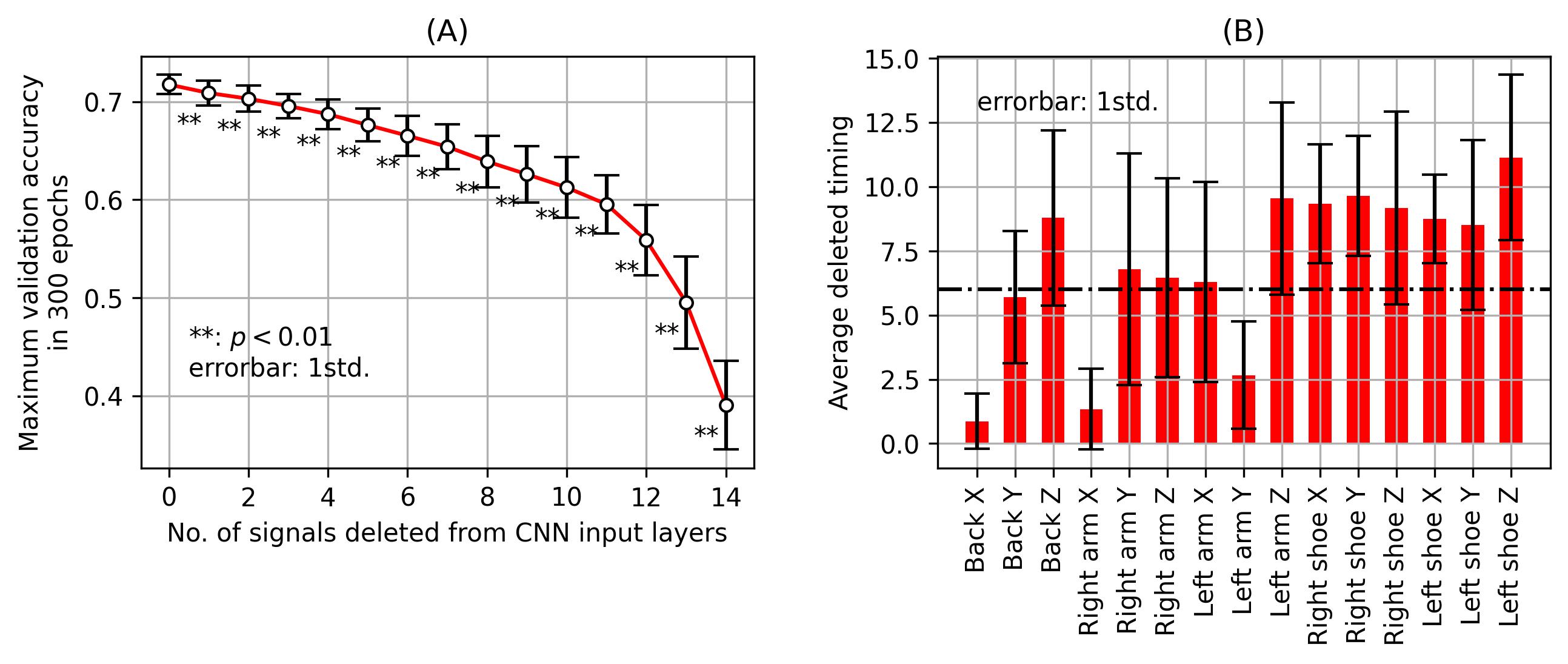}
 \caption{
(A) Maximum validation accuracy in 300 epochs when the most important signal $s_\m{max}$ is gradually removed and CNNs are re-learned.
 (B) Average removed timings of each signal. 
 In contrast to that shown in Figure \ref{f2}, the most important signals are removed.}
 \label{f3}
\end{figure*}

\section{Experiment 1: Relationship between validation accuracy and the number of deleted signals}
\subsection{Objective and outline}
Herein, we examine the reliability of $s_\m{min,\ max}$ defined in Equation \ref{eq_imp}.
We expect to maintain estimation accuracy even if the signal $s_\m{min}$ is removed because it is the most nonimportant signal.
By contrast, because signal $s_\m{max}$ is the most important signal, the estimation accuracy may decrease by removing it.
To verify the aforementioned hypothesis, we gradually removed signal $s_\m{min,\ max}$ using the processes of lines 2 to 7 of Algorithm \ref{alg1}.
In addition, we repeatedly developed CNNs and recorded their validation accuracies.
We performed verification using a total of 100 seeds because CNN depends on randomness.
The adopted layer structure of the CNN is explained in the Appendix section; the number of epochs was 300.

\subsection{Result and discussion}
First, we indicate the validation accuracies when the most nonimportant signal, $s_\m{min}$, is gradually removed, as shown in Figure \ref{f2} (A).
From left to right in Figure \ref{f2} (A), the number of deleted signals increases, that is, the number of signals in the CNN input layer decreases.
The leftmost result is obtained using all signals, and the rightmost result is obtained by using only one signal as the CNN input layer.
The results show that even if six signals were deleted, the validation accuracy did not decrease.
Moreover, by removing seven or more signals, the accuracy decreases significantly.
In this case, although CNNs estimate class labels from 15 signals in the set $\bm{S}$, six signals appear unnecessary.

We calculated the average removed timings of the 15 signals to determine the unnecessary signals.
The result is shown in Figure \ref{f2} (b).
This means that the lower the value, the earlier the signal is removed in the procedure of Algorithm \ref{alg1}.
The results indicate that the signals of the right and left shoes are removed earlier than those of the other sensors.
By contrast, some signals from the back, right arm, and left arm did not disappear early.
Therefore, we can regard shoe signals as unimportant for classification.
In addition, shoe sensors seem important for walk motion classification.
However, even if shoe sensors disappear, we consider that the back sensor attached to centroids of the human body contributes to the classification of walk motions because these motions are periodic.

Next, we indicate the validation accuracies when the most important signal, $s_\m{max}$, is gradually removed in Figure \ref{f3} (a).
We can verify that the validation accuracy statistically decreases by removing one of the most important signals.
Therefore, we consider that removing the most important signal, $s_\m{max}$ leads to a worse accuracy.
The average timing of the signal removal is shown in Figure \ref{f3} (b).
From this figure, we confirm that the signals of back X, right arm X, and left arm Y are removed early.
Moreover, we confirm that shoe sensor signals are not removed early.
These tendencies are in contrast compared with the case of removing the most nonimportant signal $s_\m{min}$.

Clearly, (1) even if we remove the most nonimportant signal $s_\m{min}$, the estimation performance tends to remain, and (2) when we remove the most important signal $s_\m{max}$, the performance tends to decrease.
Therefore, we regard the signal importance $I^s(\bm{X})$ defined in Equation \ref{eq_imp2} as reliable. 

\section{Experiment 2: Effectiveness of FG-SSA on generalization scores} \label{sec4}
\subsection{Objective and outline}
Herein, we confirm the effect of FG-SSA indicated in Algorithm \ref{alg1} on the generalization performance of the classification.
Therefore, we develop the following three conditions for CNNs.
\begin{itemize}
\item Condition A: CNN using all signals, i.e., FG-SSA is not used.
\item Condition B: CNN for applied FG-SSA of $\gamma = n_\m{s}$.
\item Condition C: CNN for applied FG-SSA of $\gamma = 9$.
\end{itemize}
Condition A means that the CNN does not remove signals but uses all signal sets $\bm{S}$.
Condition B refers to the CNN use the signal subset $\bm{S}_\m{use}$ obtained by the FG-SSA, given $n_\m{s}$ as the constraint parameter $\gamma$.
Under this condition, when maximum validation accuracy is achieved using all signal sets $\bm{S}$, we allow the algorithm to return $\bm{S}$ as $\bm{S}_\m{use}$. 
In other words, a case wherein no signals are removed can occur.
Condition C implies that the number of adopted signals is nine or less for a CNN input layer, that is, $|\bm{S}_\m{use}| \le 9$.
In other words, six or more signals were deleted, because the initial number of signals was 15.
The value of hyper-parameter $\gamma = 9$ was determined by referring to the result shown in Figure \ref{f2} (A).

We developed CNNs using the signal set, which was determined by FG-SSA.
Moreover, the optimal epochs leading to a maximum validation accuracy were adopted.
The search range of epochs was from 1 to 300.
Subsequently, the generalization performance is measured using the test dataset.
The test dataset was used only at this time, that is, it was not used for the parameter search.
To remove the randomness effect, we developed CNNs for Conditions A, B, and C with a total of 100 seeds.

\begin{figure*}[t]
 \centering
 \includegraphics[scale=0.6]{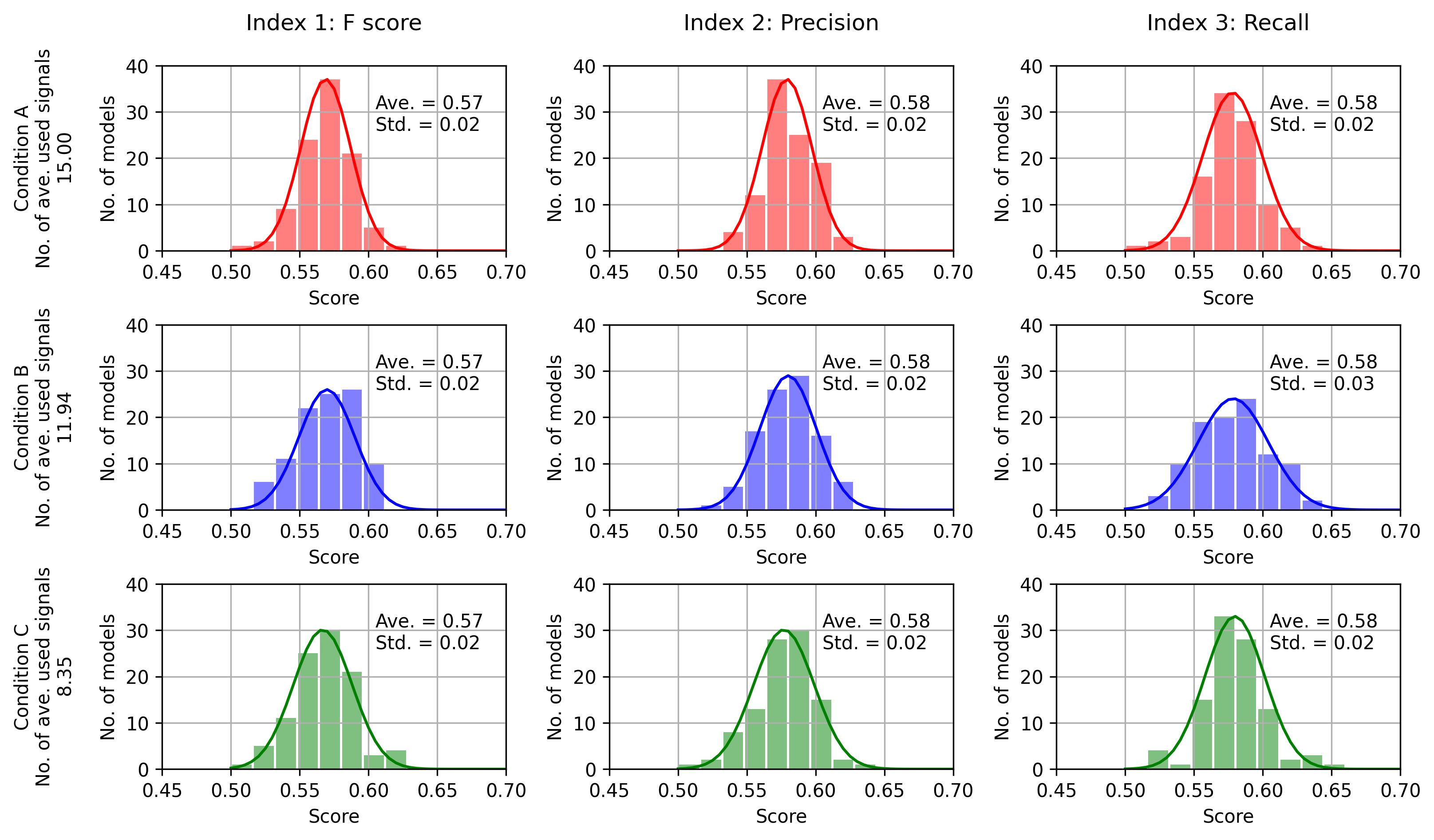}
 \caption{Histograms of estimation score using the test dataset (total 100 seeds). 
 The macro averages of 10 classes of F score, precision, and recall from left to right, and Conditions A, B, and C from top to bottom.}
 \label{f4}
\end{figure*}

\begin{figure*}[t]
 \centering
 \includegraphics[scale=0.42]{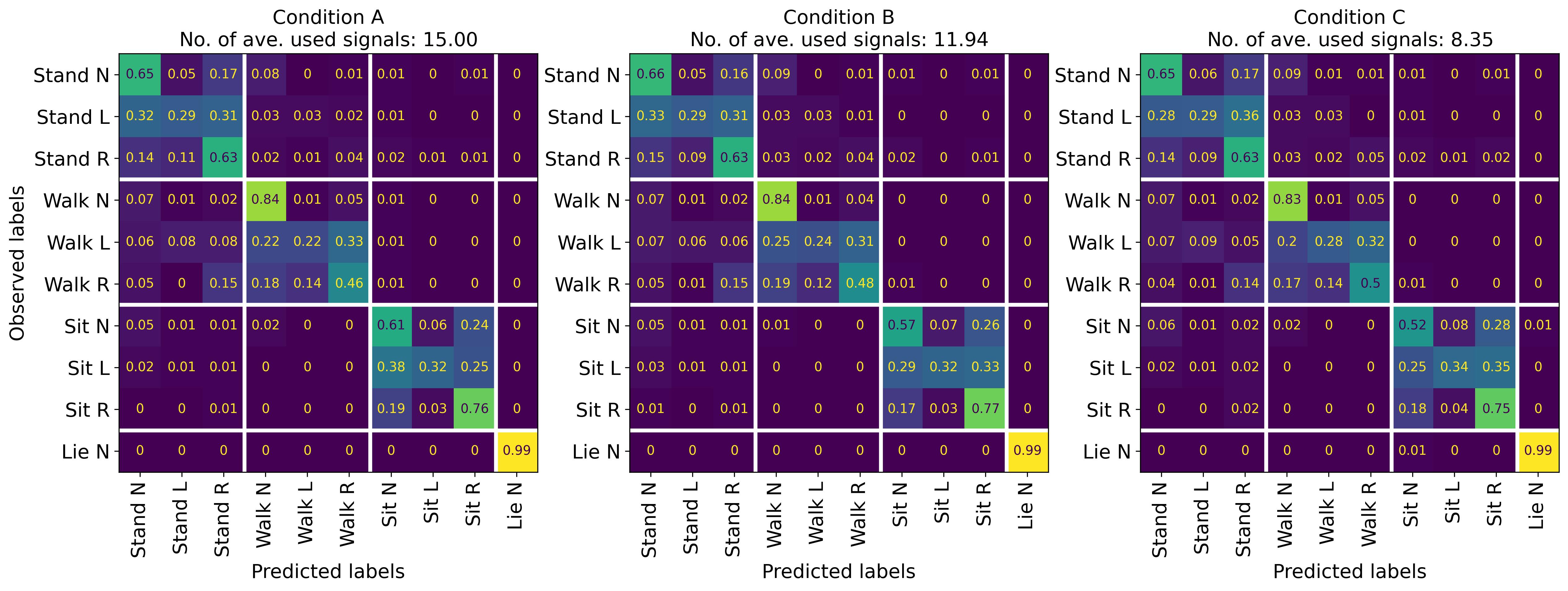}
 \caption{Confusion matrices of the conditions A, B, and C using the test dataset. Values are averaged by results of 100 seeds. The values are standardized wherein the summation in each row is one.}
 \label{f5}
\end{figure*}

\subsection{Result and discussion}
By developing CNNs under conditions A, B, and C on a total of 100 seeds, the average number of signals used was 15.00, 11.94, and 8.35, respectively.
In other words, an average of 3.06 and 6.65 signals were removed by FG-SSA in the cases of Condition B and C, respectively.
The generalization performance (F score, precision, and recall) measured by the test dataset for each condition is shown in Figure \ref{f4}.
These are histograms of 100 CNNs developed using 100 random seeds.

The results indicated that the generalization scores were nearly the same for conditions A, B, and C.
Moreover, $p$ values of the two-sided $t$-test were not statistically significant.
Next, we show the confusion matrices obtained by CNNs for each condition in Figure \ref{f5}.
The values were averaged by 100 seed results and standardized, where the summation in each row was 1.
Consequently, the confusion matrices were nearly the same.
Although the proposed algorithm removed some signals, generalization errors did not increase.
Therefore, we consider FG-SSA has the effect of finding and removing signals that are not important for CNN-based classification.

From another viewpoint, the number of correct classifications of left hand motions (Stand L, Walk L, and Sit L) was small in all conditions.
We adopted weighted cross-entropy-based learning for CNNs because the original data size of the left hand motions is small, as indicated in Table \ref{tab1}.
However, we consider that an appropriate classification cannot be performed because the data diversity of these motions is low.
Although we believe that nearly all subjects are right-handed, the explanation of subjects' dominant arm in the OPPORTUNITY dataset is insufficient to the best of our knowledge.

\section{Conclusion, limitations, and future works}
In this paper, we describe the following two topics to find and remove nonimportant signals for CNN-based classification.
\begin{itemize}
\item (1) The signals importance indices SIM $\bm{I}_\m{mat}(\bm{X})$ and SIV $\bm{I}_\m{vec}(\bm{X})$ are explained in Subsection \ref{sec2.4}.
\item (2) The algorithm of the linear complexity $\mathcal{O}(n_\m{s})$ for obtaining the signals subset $\bm{S}_\m{use}$ from the initial signals set $\bm{S}$ by finding and removing nonimportant signals (see Algorithm \ref{alg1}).
\end{itemize}

Although the proposed algorithm performed well in the case of the OPPORTUNITY dataset, it had some limitations. This is explained in the following sections. Future work will confirm these findings.
\begin{itemize}
\item (A) The results described in this paper is obtained from limited cases. Although we assume that the algorithm can be applied to other than acceleration signals, e.g., EEG, ECG, and others, it is not validated. 
\item (B) In this experiment, the initial signals set size was 15. The number of signals may be much higher depending on some situations. It is important to determine the relationship between the number of nonimportant signals and effectiveness of FG-SSA.
\item (C) We assume that the algorithm may extract the signals subset weighted specific target class by providing class weights to the signal importance $I^\m{s}(\bm{X})$ defined in Equation \ref{eq_imp2}. 
In other words, we denote the importance of signal $s$ weighted classes as
\begin{align}
I^s(\bm{X}; \bm{w}) = \frac{1}{|\bm{C}|}\sum_{c^\p \in \bm{C}} w^{c^\p} I^{s, c^\p}_\m{mat}(\bm{X}),
\end{align}
where $w^{c^\p}$ denotes the weight of class $c^\p$ and $\bm{w}$ denotes the class weight vector, defined as follows:
\begin{align}
\bm{w} &= [w^{c^\p}] \in \mathbb{R}^{|\bm{C}|}, \text{ s.t., } \sum_{c^\p \in \bm{C}} w^{c^\p} = 1.
\end{align}
We note that the algorithm based on $I^s(\bm{X}; \bm{w})$ can return the signals subset weighted specific classes.
However, this effect is not confirmed in this paper.
This will be studied in the future.
\end{itemize}

\section*{Appendix: Layers structure of CNNs}
The CNNs layer structure is illustrated in Figure \ref{f0}.
First, an input layer exists for inputting $w \times n_\m{s}$ size data ($w$: window length, $n_\m{s}$: the number of signals). 
Then, $n_\m{c}=3$ convolution layers exist to generate $n_\m{f}$ feature maps using kernel filters of convolution size $s_\m{c}$.
The convolutions are only in the direction of time.
Next, the generated feature maps are transformed into a vector form by the flattened layer.
Vector dimensions are gradually reduced to 200, 100, and 50.
Finally, an output vector $\bm{y}$ of 10 dimensions is generated, and classification is performed.
The total number of CNN layers is nine (one input layer, three convolution layers, four dense layers, and one output layer).
The activation function of each layer is ReLu.

We performed a hyperparameter search for
convolution size $s_\m{c}$ and 
the number of kernel filters $n_\m{f}$ and
learning rate $r$.
In particular, CNNs were trained using the training and validation datasets described in Subsection \ref{se21} and 300 epochs.
We adopted a weighted cross-entropy-based loss function based on the inverse values of class sample sizes because the sample sizes of each class are imbalanced, as listed in Table \ref{tab1}.

The parameter candidates are as follows:
\begin{align}
&s_\m{c} \in \{5, 10, 15 \}, \nonumber \\
&n_\m{f} \in \{5, 10\}, \nonumber \\
&r \in \{10^{-3}, 10^{-4}\}.
\end{align}
The combination of hyperparameters leading to the maximum validation accuracy was $(s_\m{c}, n_\m{f}, r) = (10, 10, 10^{-4})$ (accuracy: 0.726). 
Therefore, we adopted these parameters because all CNNs appeared in this study.
Although CNNs have many other hyperparameters, we refrained from excessive tuning because the main purpose of this study is to provide and confirm the signal-selection algorithm.

\nocite{*}
\bibliographystyle{IEEEtran}
\bibliography{ieee_refs}

\end{document}